\documentclass[epjCONF, onecolumn]{svjour}
\usepackage{graphics}
\usepackage[varg]{txfonts} 
\usepackage[latin1]{inputenc}
\session-title{Assembling the puzzle of the Milky Way}

\newcommand {\hi} {{\rm H}\,{\small\rm I}}

\newcommand {\kms} {\,{\rm km\,s}^{-1}}

\newcommand {\kpc} {\,{\rm kpc}}

\newcommand {\de}{^{\circ}}

\newcommand{\gsim}{\lower.7ex\hbox{$\;\stackrel{\textstyle>}{\sim}\;$}}
\newcommand{\lsim}{\lower.7ex\hbox{$\;\stackrel{\textstyle<}{\sim}\;$}}

\newcommand {\apj}{ApJ}

\newcommand {\mnras}{MNRAS}
\newcommand {\aap}{A\&A}
\newcommand {\pasj}{PASJ}

\newcommand {\bain}{Bulletin of the Astronomical Institutes of the Netherlands}

\begin{document}
\title{How axi-symmetric is the inner \hi\ disc of the Milky Way?}
\author{Antonino Marasco\inst{1}\fnmsep\thanks{\email{antonino.marasco2@unibo.it}} \and Filippo Fraternali\inst{1,2}}
\institute{Department of Astronomy, University of Bologna, via Ranzani 1, 40127, Bologna, Italy \and Kapteyn Astronomical Institute, Postbus 800, 9700 AV, Groningen, The Netherlands}
\abstract{
We modelled the distribution and the kinematics of \hi\ in the inner Milky Way ($R\!<\!R_\odot$) at latitude $b\!=\!0\de$ assuming axi-symmetry. 
We fitted the line profiles of the LAB 21-cm survey using an iterative approach based on the tangent-point method. 
The resulting model reproduces the \hi\ data remarkably well, with significant differences arising only for $R\lsim 2\kpc$. 
This suggests that, despite the presence of a barred potential, the neutral gas in the inner Milky Way is distributed in a fairly axi-symmetric disc.}
\maketitle
\section{The method}
The rotation curve of the inner Milky Way is classically derived using the so-called \emph{tangent-point} method, which is based on extracting the terminal-velocity from the \hi\ line profiles \cite{Shane66,Malhotra95}. This is usually done by assuming the velocity dispersion or the density of the gas to be constant for a wide range of radial velocities \cite{Celnik79,Rohlfs87}.
We developed an iterative method to  fit the whole \hi\ line profiles of the LAB 21-cm survery \cite{Kalberla05} at latitude $b=0\de$, assuming that the neutral gas in the inner Galaxy is distributed in rings in circular rotation, centered at the Galactic centre.
\begin{figure}[h]
\centering
\resizebox{0.9\columnwidth}{!}{
\includegraphics{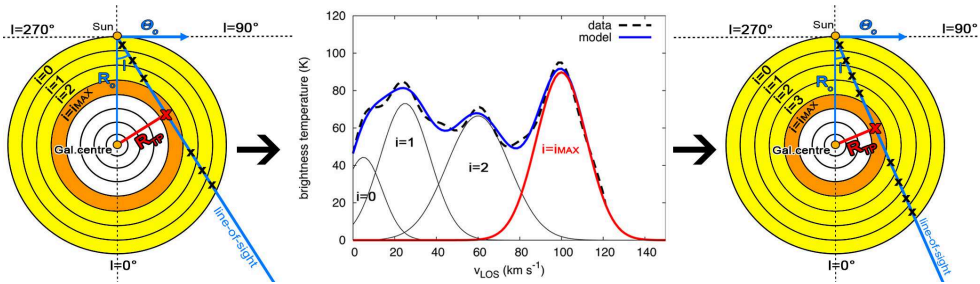}}
\caption{Scheme of the recursive tangent-point method. The \emph{first panel} sketches the inner Milky Way divided into rings. The \hi\ line profile at longitude $l$ is given by the sum of the contributions of all rings with $0\le i\le i_{\rm max}$ (\emph{second panel}), the parameters of the last ring can be found by fitting the last gaussian of the model profile (thick red curve) to the data (dashed curve). This procedure can be iterated by decreasing $l$ (\emph{third panel}).}
\label{scheme}
\end{figure}
Our iterative method is sketched in Fig.\,\ref{scheme}. 
The generic $i$-th ring at radius $R_i$ is described by three parameters: the rotational velocity $\Theta_i$, the velocity dispersion $\sigma_i$ and the volume density $n_i$ of the gas. 
Hence, at longitude $l$, the line profile $T_l (v)$ can be modelled by summing the contributions (assumed Gaussian) of all rings with $0\!\le\!i\!\le\!i_{\rm max}$ along the line of sight, where $R_0\!=\!R_{\odot}$ and $R_{i_{\rm max}}\!=\!R_{\odot}\sin(l)$ is the tangent-point radius $R_{\rm tp}$.
Thus $T_l (v) = 2\sum_{i=0}^{i_{\rm max}} \delta(\Theta_i, \sigma_i, n_i)\, \Delta d_i $,
where the factor $2$ stands for axi-symmetry, $\delta$ is a Gaussian function centered at $\left(\Theta_i\,\frac{R_{\odot}}{R_i}-\Theta_{\odot}\right)\sin(l)$ with dispersion $\sigma_i$ and amplitude $\propto n_i$, and $\Delta d_i$ is the separation along the line of sight between two consecutive rings. 
The line profile is then corrected for the optical thickness by assuming the gas isothermal with a temperature of $152$ K (maximum brightness temperature in the LAB).
We applied this method to determine the values of the parameters for all rings recursively, starting from $R=R_{\odot}$ (at $l\!=\!90\de$ for the receding region, or $l=270\de$ for the approaching one) down to the Galactic centre (at $l=0\de$). 
Note that for each new $l$ we fix the values of $\Theta_i$, $\sigma_i$ and $n_i$ for all the rings with $i\!<\!i_{\rm max}$ determined with the previous iterations and fit only the values at $i\!=\!i_{\rm max}$. 
We assume $R_{\odot}=8.3\kpc$ and $\Theta_{\odot}=240\kms$ \cite{McMillan11}.
\section{Results} \label{results}
\begin{figure*}[ht]
\centering
\resizebox{0.55\columnwidth}{!}{
\includegraphics{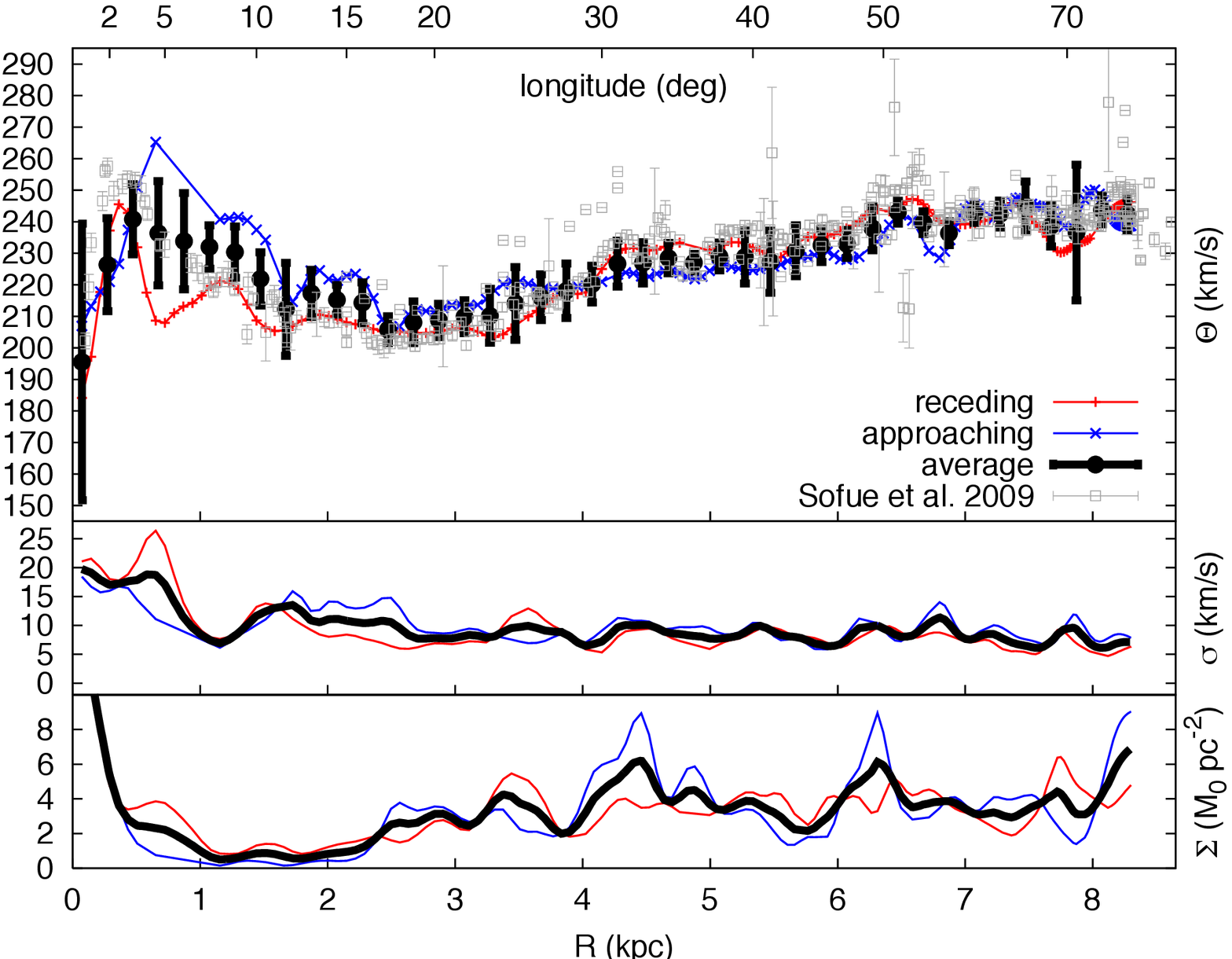}}
\resizebox{0.44\columnwidth}{!}{
\includegraphics{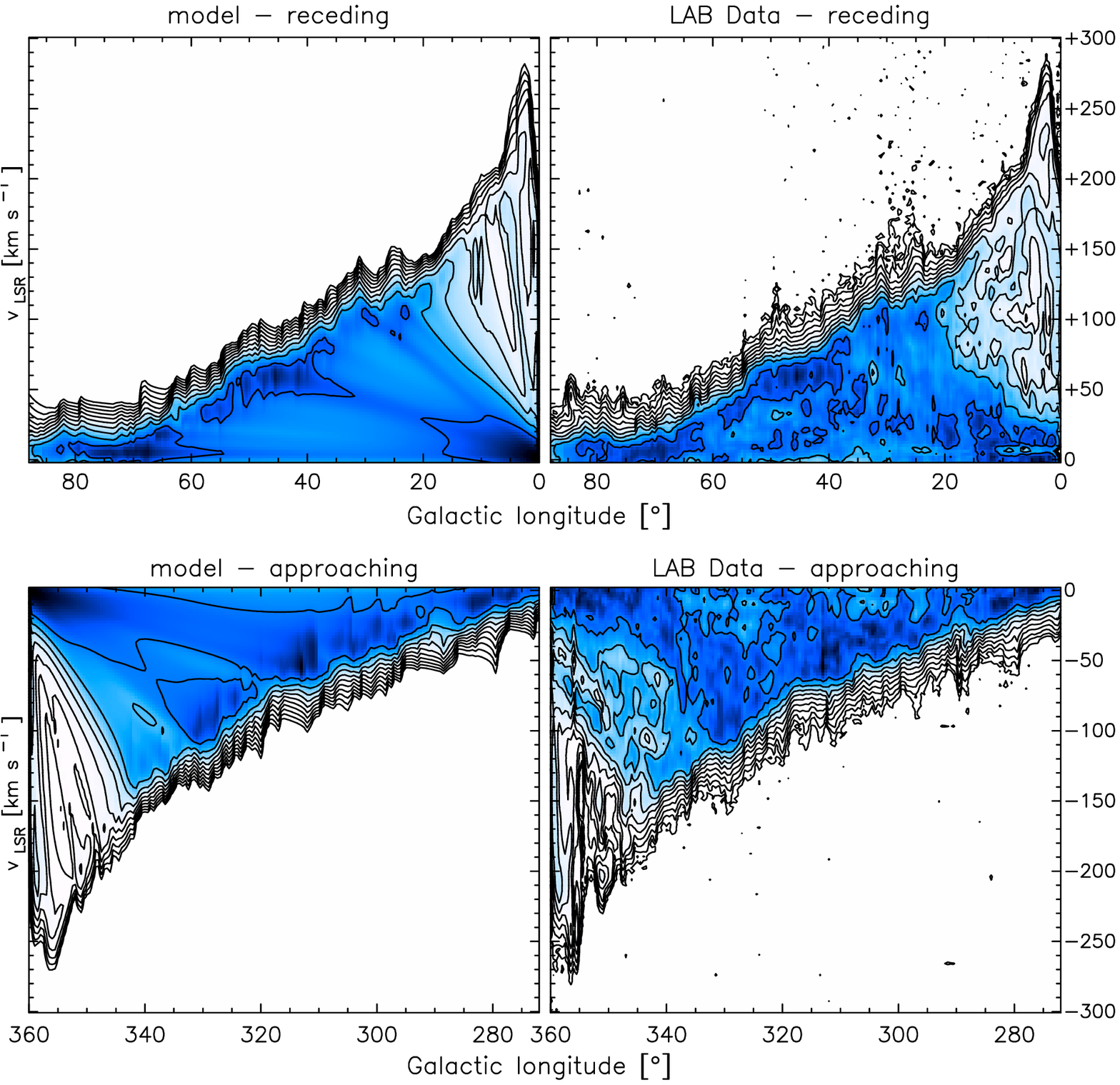}}
\caption{Results of our axi-symmetric model for the inner Milky-Way disc. \emph{Left panels}: rotation curve, dispersion and density profiles for the approching (blue) and receding (red) sides.
\emph{Righ panels}: Comparison between our model and the LAB data, $l-v$ slices at $b=0\de$. Contour levels range from $0.16$ K to $81.92 K$ scaling by a factor 2.}
\label{models}
\end{figure*}
Figure \ref{models} (left panels) shows $\Theta(R)$, $\sigma(R)$ and the \hi\ surface density $\Sigma(R)$ as derived with our method. 
Error bars on the average rotation curve take into account the differences between the approaching and the receding velocities, which are significant ($\gsim20\kms$) only for $R\lsim 2\kpc$ and are likely due to non-circular motions in the region of the bar. 
Our average rotation curve is consistent with the classical one \cite{Sofue09}.
The velocity dispersion is rather high in the innermost regions decreasing to about $7\!-\!8\kms$ for $R\!>\!3\kpc$.
The surface density profile has been derived by assuming a constant scale-height of $150$ pc. 
The right panels of Figure \ref{models} compare the longitude-velocity ($l-v$) plots at $b=0\de$ predicted by our model with those of the data. 
It is remarkable that a fully axi-symmetric description reproduces the line profile in such detail including the emission far from the terminal velocities.
This seems to indicate that the neutral gas in the inner Galaxy is not too much affected by the non-axisymmetry of the potential and its kinematics is dominated by rotation.

\end{document}